# Interferometric characterization of pulse front tilt of spatiotemporally focused femtosecond laser pulses for three-dimensional micromachining application

Zhaohui Wang,[1,2] Fei He,[1] Jielei Ni,[1] Chenrui Jing,[1,2] Hongqiang Xie,[1,2] Bin Zeng,[1] Wei Chu,[1] Lingling Qiao,[1] Ya Cheng,[1,*]

[1] *State Key Laboratory of High Field Laser Physics, Shanghai Institute of Optics and Fine Mechanics, Chinese Academy of Sciences, P.O. Box 800-211, Shanghai 201800, China*
[2] *University of Chinese Academy of Science, Beijing 100049, China*
*ya.cheng@siom.ac.cn

**Abstract:** We report on an experimental measurement of the pulse front tilt (PFT) of spatiotemporally focused femtosecond laser pulses in the focal plane in both air and bulk transparent materials, which is achieved by examination of the interference pattern between the spatiotemporally focused pulse and a conventional focused reference pulse as a function of time delay between the two pulses. Our simulation results agree well with the experimental observations.

**References and links**

1. R. R. Gattass, and E. Mazur, "Femtosecond laser micromachining in transparent materials," Nature Photon. **2**(4), 219-225 (2008).
2. K. Sugioka, and Y. Cheng, "Ultrafast lasers—reliable tools for advanced materials processing," Light: Sci. Appl. **3**(4), e149 (2014).
3. R. Osellame, H. J. Hoekstra, G. Cerullo, and M. Pollnau, "Femtosecond laser microstructuring: an enabling tool for optofluidic lab‐on‐chips," Laser Photon. Rev. **5**(3), 442-463 (2011).
4. M. Beresna, M. Gecevičius, and P. G. Kazansky, "Ultrafast laser direct writing and nanostructuring in transparent materials," Adv. Opt. Photon. **6**(3), 293-339(2014).
5. K. Itoh, W. Watanabe, S. Nolte, and C. Schaffer, "Ultrafast processes for bulk modification of transparent materials," MRS Bull. **31**(08), 620-625(2006).
6. M. Ams, G. D. Marshall, P. Dekker, J. A. Piper, and M. J. Withford, "Ultrafast laser written active devices," Laser Photon. Rev. **3**(6), 535-544(2009).
7. F. Chen, and J. R. Aldana. "Optical waveguides in crystalline dielectric materials produced by femtosecond‐laser micromachining,"  Laser Photon. Rev. 2014, **8**(2): 251-275.
8. S. Kawata, H. B. Sun, T. Tanaka, and K. Takada, "Finer features for functional microdevices," Nature **412**(6848), 697-698 (2001).
9. S. Juodkazis, K. Nishimura, S. Tanaka, H. Misawa, E. G. Gamaly, B. Luther-Davies, L. Hallo, P. Nicolai, and V. T. Tikhonchuk, "Laser-induced microexplosion confined in the bulk of a sapphire crystal: evidence of multimegabar pressures,"  Phys. Rev. Lett. **96**(16), 166101(2006).
10. Y. Liao, Y. Cheng, C. Liu, J. Song, F. He, Y. Shen, D. Chen, Z. Xu, Z. Fan, X. Wei, K. Sugioka, and K. Midorikawa, "Direct laser writing of sub-50 nm nanofluidic channels buried in glass for three-dimensional micro-nanofluidic integration," Lab Chip **13**(8), 1626-1631 (2013).
11. R. Osellame, S. Taccheo, M. Marangoni, R. Ramponi, P. Laporta, D. Polli, S. D. Silvestri, and G. Cerullo, "Femtosecond writing of active optical waveguides with astigmatically shaped beams," JOSA B **20**(7), 1559-1567 (2003).
12. Y. Cheng, K. Sugioka, K. Midorikawa, M. Masuda, K. Toyoda, M. Kawachi, and K. Shihoyama, "Control of the cross-sectional shape of a hollow microchannel embedded in photostructurable glass by use of a femtosecond laser," Opt. Lett. **28**(1), 55-57 (2003).
13. K. Sugioka, Y. Cheng, K. Midorikawa, F. Takase, and H. Takai "Femtosecond laser microprocessing with three-dimensionally isotropic spatial resolution using crossed-beam irradiation," Opt. Lett. **31**, 208-210 (2006).

14. F. He, H. Xu, Y. Cheng, J. Ni, H. Xiong, Z. Xu, K. Sugioka, and K. Midorikawa, "Fabrication of microfluidic channels with a circular cross section using spatiotemporally focused femtosecond laser pulses," Opt. Lett. **35**(7), 1106-1108 (2010).
15. F. He, Y. Cheng, J. Lin, J. Ni, Z. Xu, K. Sugioka, and K. Midorikawa, "Independent control of aspect ratios in the axial and lateral cross sections of a focal spot for three-dimensional femtosecond laser micromachining," New J. Phys. **13**(8) 083014 (2011).
16. F. He, B. Zeng, W. Chu, J. Ni, K. Sugioka, Y. Cheng, and C. G. Durfee, "Characterization and control of peak intensity distribution at the focus of a spatiotemporally focused femtosecond laser beam, " Opt. Express **22**(8), 9734-9748 (2014).
17. C. G. Durfee, M. Greco, E. Block, D. Vitek, and J.A. Squier, "Intuitive analysis of space-time focusing with double-ABCD calculation," Opt. Express **20**(13), 14244-14259 (2012).
18. D. N. Vitek, E. Block, Y. Bellouard, D. E. Adams, S. Backus, D. Kleinfeld, C. G. Durfee, and J. A. Squier, "Spatio-temporally focused femtosecond laser pulses for nonreciprocal writing in optically transparent materials," Opt. Express **18**(24), 24673-24678(2010).
19. R. Kammel, R. Ackermann, J. Thomas, J. Götte, S. Skupin, A. Tünnermann, and S. Nolte, "Enhancing precision in fs-laser material processing by simultaneous spatial and temporal focusing," Light: Sci. Appl. **3**(5), e169(2014).
20. P. G. Kazansky, W. Yang, E. Bricchi, J. Bovatsek, A. Arai, Y. Shimotsuma, K. Miura, and K. Hirao, "Quill writing with ultrashort light pulses in transparent materials," Appl. Phys. Lett. **90**(15), 151120(2007).
21. P. S. Salter, and M. J. Booth, "Dynamic control of directional asymmetry observed in ultrafast laser direct writing," Appl. Phys. Lett. **101**(14), 141109(2012).
22. J. Goodman, *Introduction to Fourier optics* (Roberts & Company, Englewood, Colorado, 2005).
23. S. Hell, G. Reiner, C. Cremer, and E. H. Stelzer, "Aberrations in confocal fluorescence microscopy induced by mismatches in refractive index," J. Microsc. **169**(3), 391-405(1993).
24. A. Jesacher, and M. J. Booth, "Parallel direct laser writing in three dimensions with spatially dependent aberration correction," Opt. Express **18**(20), 21090-21099(2010).
25. B. Sun, P. S. Salter, and M. J. Booth, "Effects of aberrations in spatiotemporal focusing of ultrashort laser pulses," JOSA A **31**(4), 765-772(2014).

## 1. Introduction

Femtosecond laser pulses have been widely used for advanced materials processing on both micro and nano scales [1-7]. Once tightly focused and employed for direct writing, such intense femtosecond laser pulses intrinsically provide capabilities of three-dimensional (3D) micro and nano fabrication due to efficient confinement of the nonlinear interactions within the focal volume, enabling sub-100 nm stereolithography based on two-photon polymerization and micro and nanoscale internal modification of bulk transparent materials [8-10]. Unlike the conventional two-dimensional (2D) planar lithographic fabrication in which only the transverse fabrication resolution should be considered, the longitudinal (*i.e.*, the axial direction which is parallel to the propagation direction of the laser beam) fabrication resolution plays an equally important role in 3D femtosecond laser processing in terms of the quality of the fabricated structures. To improve the longitudinal resolution, several beam shaping techniques such as astigmatic beam shaping [11], slit beam shaping [12], crossed-beam shaping [13], and spatiotemporal beam shaping [14-19] have been demonstrated. It is noteworthy that the spatiotemporal beam shaping technique is the only technique that currently allows for achieving a 3D isotropic fabrication resolution with a single focal lens, which has been experimentally proved by fabrication of microfluidic channels of circular cross sectional shape regardless of the direction of sample translation [14]. Nevertheless, it has recently been pointed out by Vitek *et al*. that the spatiotemporally focused spot intrinsically has a tilted pulse front, leading to a nonreciprocal writing effect [18]. The nonreciprocal writing was also observed earlier by Yang *et al*. with conventionally focused femtosecond laser pulses, which gives rise to anisotropic fabrication quality depending on the direction of sample translation [20]. Currently, the mechanism behind the nonreciprocal writing effect has not been fully understood; however, its connection to the pulse front tilt (PFT) has been clearly confirmed by several recent investigations [18, 20, 21]. To gain a deep insight of the nonreciprocal writing, quantitative characterization of the PFT of femtosecond laser pulses focused into transparent media is necessary.

In this article, we report on direct measurement of the tilting angle of front-tilted pulses based on an interferometric measurement scheme. Our technique, although being simple and easy to operate, can be used for characterization of spatiotemporally focused pulses generated in either air or bulk transparent materials. We present both results obtained in air and in fused silica glass for comparison.

The concept of this technique is illustrated in Fig. 1. It is known that when focusing a femtosecond laser pulse with a conventional focusing (CF) systems (*i.e.*, an objective lens), no PFT effect will be induced. This pulse will be used as a reference pulse in our experiment. However, spatiotemporal focusing (STF) of a femtosecond laser pulse will lead to sweep of the focal spot across the focal plane at a sweeping velocity which is inversely proportional to the degree of the PFT. The sweeping direction depends on the direction of the spatial chirp of femtosecond laser pulses before they enter the objective lens. Thus, by intentionally varying the time delay between the reference pulse and the spatiotemporally focused femtosecond pulses, the intersection of two pulses will induce a line-shaped interference pattern which shifts linearly with the time delay between the two pulses, as illustrated by Figs. 1(a)-(c). The line-shaped interference pattern formed in the focal plane of the lens can then be directly imaged onto a charge coupled device (CCD) using a second objective lens. By plotting the shift of the interference fringe as a function of the time delay between the two pulses, the tilting angle of the spatiotemporally focused pulse can be retrieved.

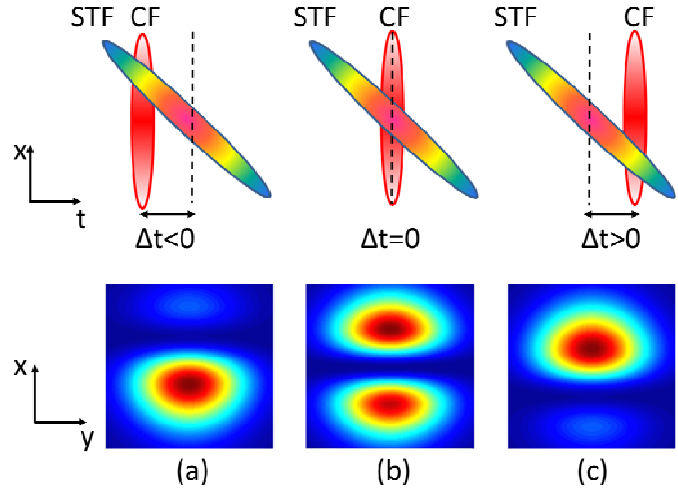

Fig.1 Schematic illustration of the concept of interferometric characterization of pulse front tilt of spatiotemporally focused femtosecond laser pulses.

## 2. Experimental setup

Figure 2 schematically illustrates the experimental setup. The femtosecond laser system (Legend-Elite, Coherent Inc.) consists of a Ti:sapphire laser oscillator and amplifier, and a grating-based stretcher and compressor, which delivers ~6 mJ, ~40 fs pulses centered at 800 nm wavelength at 1 kHz repetition rate. Before the amplified laser beam was recompressed, it was split into two using a 1:1 beam splitter. One beam, which was used to produce the spatiotemporally focused pulse, was spatially dispersed along the *x* direction by a pair of 1500 lines/mm gratings (blazing at ~53°) which were arranged to be parallel to each other. The distance between the two gratings was adjusted to be ~780 mm to compensate for the temporal dispersion of pulses. After being dispersed by the grating pair, the laser beam was measured to be ~40 mm (1/e$^2$) along the *x*-axis and ~10 mm (1/e$^2$) along the *y*-axis, as indicated in Fig. 2. The other beam was compressed by passing through the compressor to generate the reference beam. The spatially dispersed beam and the reference beam were then collinearly recombined using another 1:1 beam splitter. A delay line was inserted into the optical path of the reference beam to adjust the time delay between the two beams. These two beams were then focused in either air or fused silica using a plane-convex lens with a focal length of *f* = 750 mm and a diameter of 100 mm. The interference patterns were imaged with an objective lens (20×, NA=0.4) on the CCD camera (Wincamd-ucd23) with an integration time of 5 ms. The average power of each beam was individually controlled using a variable neutral density (ND) filter for optimizing the contrast of the interference fringes.

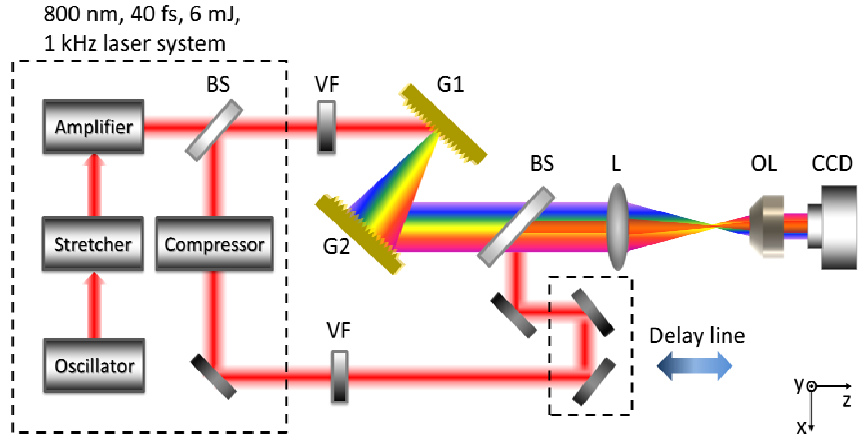

Fig. 2 Schematic of the experimental setup. BS: beam splitter, G1-2: gratings (1500l/mm), L: lens (f=75cm), OL: objective lens (20×, NA=0.40), VF: variable neutral density filter.

### 3. Experimental results

First, we present our measurement results obtained in air in Fig. 3. To perform accurate measurement of the sweeping velocity of the spatiotemporally focused spot, a linear propagation condition is preferred. Thus, the laser pulse energy was set at ~0.3 μJ, which is sufficiently low to avoid any nonlinear propagation effects in air or in fused silica under our focusing condition. Figs. 3(a) and (b) present the images of the focal spots generated with the conventional focusing (CF) and the spatiotemporal focusing (STF) systems on the focal plane captured by the CCD camera, respectively. Both the images show circular focal spots with a Gaussian-like profile, as the images were captured with an integration time of 5 ms, which is too long to resolve the dynamic sweep process of the focal spot for the spatiotemporally focused pulses. In Figs. 3(c)-(g), the images formed by simultaneously focusing the two pulses with both the CF and STF systems are presented, in which the time delays between the two pulses are -66.8 fs, -33.4 fs, 0 fs, 33.4 fs, 66.8 fs, respectively. Here, the positive time delay means that the pulse focused by the STF system arrives earlier than the pulse focused by the CF system, and vice versa. One can see that in these images, interference fringes have been formed due to the coherent superposition between the two pulses. Due to the fluctuation in the lengths of the optical paths for the two pulses, the interference between the two pulses can be either destructive or constructive (we assume that the phase difference between the two pulses is stable during the period of each measurement because of the short integration time of 5 ms). The sweeping velocity of the interference fringe indicated by Figs. 3(c)-(g) directly maps out the sweeping velocity of the front-tilted focal spot in the focal plane, *i.e.*, the degree of PFT.

We now describe how we extract the interference fringes from the images in Figs. 3(c)-(g) by which the center of the fringes can be unambiguously determined. To remove the background signals which can be attributed to the incoherent superposition between the two pulses (*i.e.*, the coherent superposition between the two pulses occurs only in the intersection of the two pulses as illustrated in Fig. 1), we simply subtract the images of Figs. 3(a) and (b) from each image in Figs. 3(c)-(g). The image processing may generate negative values for some pixels particularly when the destructive interference occurs between the two pulses. Therefore, after the subtraction, the value of each pixel was squared. . The reproduced images in Figs. 3(h)-(l) are generated by performing the above-mentioned image processing on Figs. 3(c)-(g), respectively. In Figs. 3(h)-(l), the center of interference fringes can be rather precisely determined by looking for the locations of the highest intensities in the fringes, as indicated by the white dots. By plotting the central locations of the interference fringes as a function of time delays between the pulses generated with CF and STF systems in Fig. 3(m), one can easily fit the data with a linear function whose slope provides the sweep velocity of the focal spot on the focal plane. The measured result of the PFT is inversely proportional to the sweep velocity, which is calculated to be -2.95 fs/μm.

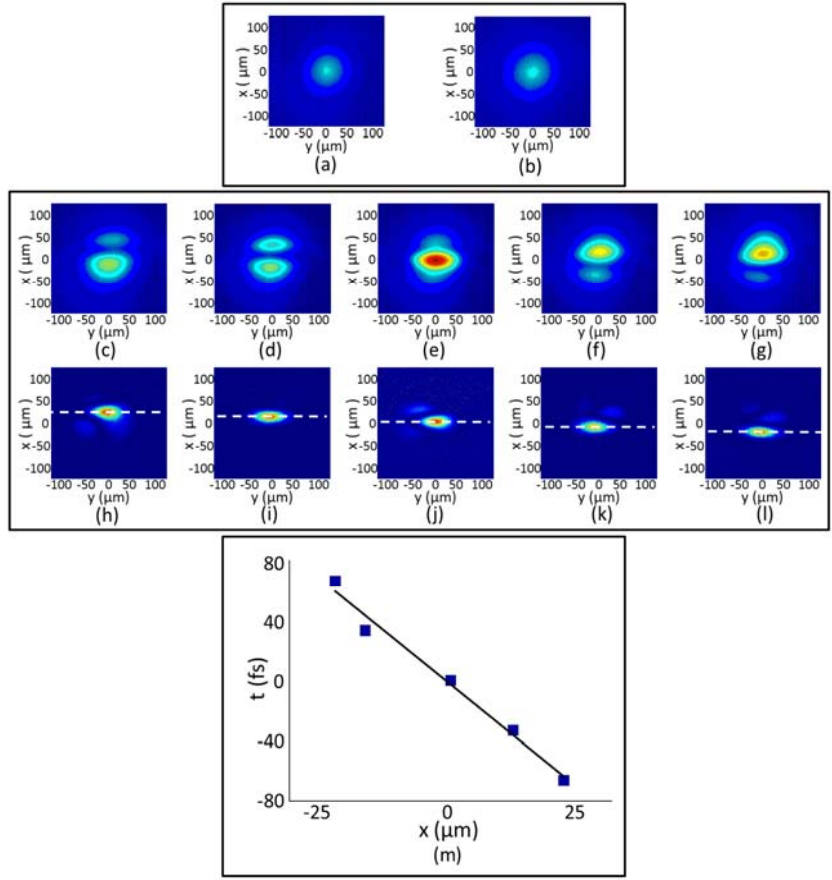

Fig. 3 Intensity distribution of the focal spot of the pulses focused with (a) STF, and (b) CF systems in the focal plane in air. Intensity distributions of the focal spots formed in air by superposition of the two beams are shown in (c)-(g). The relative time delays of the two pulses in (c)-(g) are -66.8 fs, -33.4 fs, 0 fs, 33.4 fs, 66.8 fs, respectively. (h)-(l) are the interference fringes extracted from (c)-(g), respectively. (m) displays the linear fitting of the position of the interference fringe as a function of the relative time delay.

Likewise, the same measurement can be carried out by focusing the femtosecond laser pulses in bulk transparent materials such as glass or crystals, which are directly associated with the femtosecond laser 3D micromachining application. Figures 4(a) and (b) display the CCD-captured images of the focal spots generated in a bulk fused silica sample with a thickness of 15 mm with the CF and STF systems, respectively. The depth of focus (DOF) in the glass was measured to be ~13.6 mm in this experiment. In Figs. 4(c)-(g), the images formed by simultaneously focusing the two pulses with both the CF and STF systems are presented, in which the time delays between the two pulses are again -66.8 fs, -33.4 fs, 0 fs, 33.4 fs, 66.8 fs, respectively. The magnification of the imaging system was determined by observing a microstructure with a known size behind a piece of glass with a thickness of 1.4 mm, which is the same as the distance from the focal spot in the glass to the rear surface of the glass. Therefore, the transverse size of the focal spot can be obtained from the images recorded by the CCD camera. Using the same image processing method, again we can generate a set of images in which the interference fringes in the intersection areas of the two pulses can be extracted whereas the incoherent superposition signals are removed, as shown in Figs. 4(h)-(l). The central locations of the interference fringes are then plotted in Fig. 4(m), by which the degree of the PFT is calculated to be -3.007 fs/μm. As we will show later, our measurements are consistent with the theoretical analysis.

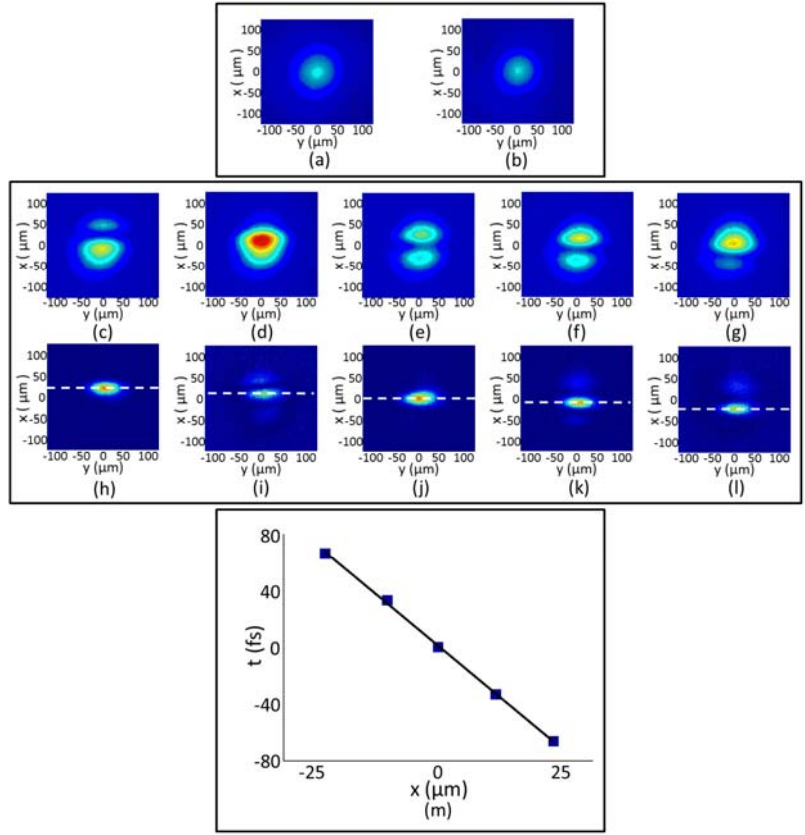

Fig. 4 Intensity distribution of the focal spot of the pulses focused with (a) STF, and (b) CF systems in the focal plane in fused silica. Intensity distributions of the focal spots formed in glass by superposition of the two beams are shown in (c)-(g). The relative time delays of the two pulses in (c)-(g) are -66.8 fs, -33.4 fs, 0 fs, 33.4 fs, 66.8 fs, respectively. (h)-(l) are the interference fringes extracted from (c)-(g), respectively. (m) displays the linear fitting of the position of the interference fringe as a function of the relative time delay.

## 4. Theoretical analysis and discussion

The measured sweeping velocity of focal spot needs to be examined by theoretical simulations using the same parameters as the experimental ones. Under the paraxial approximation, the light field in the back focal plane of the lens $E_2(x,y,\omega)$ can be obtained by performing a 2D Fourier transformation on the light field in the front focal plane $E_1(x,y,\omega)$ [22],

$$E_2(x,y,\omega) = \frac{\exp(2ikf)}{i\lambda f} \int\int_{-\infty}^{\infty} E_1(\xi,\eta,\omega) \exp\left[-\frac{ik}{f}(x\xi + y\eta)\right] d\xi d\eta, \quad (1)$$

where $\lambda$ is the wavelength, $k$ the wave vector, $f$ the focal length of the lens, and $x$, $y$ are the coordinates as shown in Fig. (2). The light field in the time domain can be obtained by performing an inverse Fourier transform on $E_2$, and the intensity distribution imaged onto the CCD camera $I(x,y)$ can be calculated from the integration of the intensity distribution over the time:

$$I(x,y) = \int_{-\infty}^{\infty} I(x,y,t) dt = \int_{-\infty}^{\infty} \left|\mathcal{F}^{-1}\left[E_2(x,y,\omega)\right]\right|^2 dt. \quad (2)$$

First, we consider the situation when the pulses are focused in air. In the situation of spatiotemporal focusing, the normalized light field of a spatially dispersed pulse $E_{STF}$ in the front focal plane can be expressed as:

$$E_{STF}(x,y,\omega) = E_0 \exp\left[-\frac{(\omega-\omega_0)^2}{\Delta\omega^2}\right] \exp\left\{-\frac{\left[x-\alpha(\omega-\omega_0)\right]^2 + y^2}{w_{in}^2}\right\}, \quad (3)$$

where $E_0$ is the constant field amplitude, $\omega_0$ the carrier frequency, $\Delta\omega$ the bandwidth (1/e$^2$ half width) of the pulses, and $w_{in}$ the initial beam waist (1/e$^2$). $\alpha(\omega-\omega_0)$ is the linear shift of each spectral component of the spatially dispersed pulse. On the other hand, the light field of a conventional femtosecond pulse $E_{CF}(x,y,\omega)$ (i.e., without spatial chirp) can be obtained by substituting $\alpha=0$ into Eq. (3),

$$E_{CF}(x,y,\omega) = E_0 \exp\left[-\frac{(\omega-\omega_0)^2}{\Delta\omega^2}\right] \exp\left\{-\frac{x^2 + y^2}{w_{in}^2}\right\}. \quad (4)$$

The intensity distribution in the back focal plane of each single beam can be obtained by substituting $E_1(x,y,\omega) = E_{STF}(x,y,\omega)$ and $E_1(x,y,\omega) = E_{CF}(x,y,\omega)$ into Eqs. (1)-(2) respectively, and the calculated results are presented in Figs. 5(a)-(b). The images in Figs. 5(a) and (b) show round-shaped focal spots with a Gaussian profile as observed in the experiment (Figs. 3(a) and (b)).

To obtain the intensity distribution of the focal spot formed by the superposed pulses in the back focal plane, the light field in the front focal plane is expressed as:

$$E_1(x,y,\omega) = E_{STF}(x,y,\omega) + E_{CF}(x,y,\omega) \exp\left[i(\omega t_d + \phi_0)\right], \quad (5)$$

where $t_d$ the relative time delay of the two pulses. Particularly, $\phi_0$ is the initial phase difference between the two pulses. The phase $\phi_0$ is artificially added into Eq. (5), because in our experiment, there is a fluctuation in the lengths of optical paths. Determination of the value of $\phi_0$ in each figure is achieved by comparing the simulation result and the experimental one until the best match between them has been achieved. By substituting Eq. (5) into Eqs. (1)-(2), we theoretically calculate the interference patterns between the two pulses in the back focal plane. The results are presented in Figs. 5(c)-(g), where the relative time delays are -66.8 fs, -33.4 fs, 0 fs, 33.4 fs, 66.8 fs, respectively. It can be seen in Fig. 5 that our simulation results reasonably reproduce the experimentally measured results in Figs. 3 (c)-(g).

Again, one can see that the interference fringe shifts linearly as a function of the relative time delay, which results from the PFT effect. The simulation result of PFT is presented in Fig. 5(h). The slope of the tilt angle of intensity distribution in $x$-$t$ domain shows that the degree of PFT is ~3.06 fs/μm, which is in good agreement with our experimental measurement with a slight difference of only ~4%.

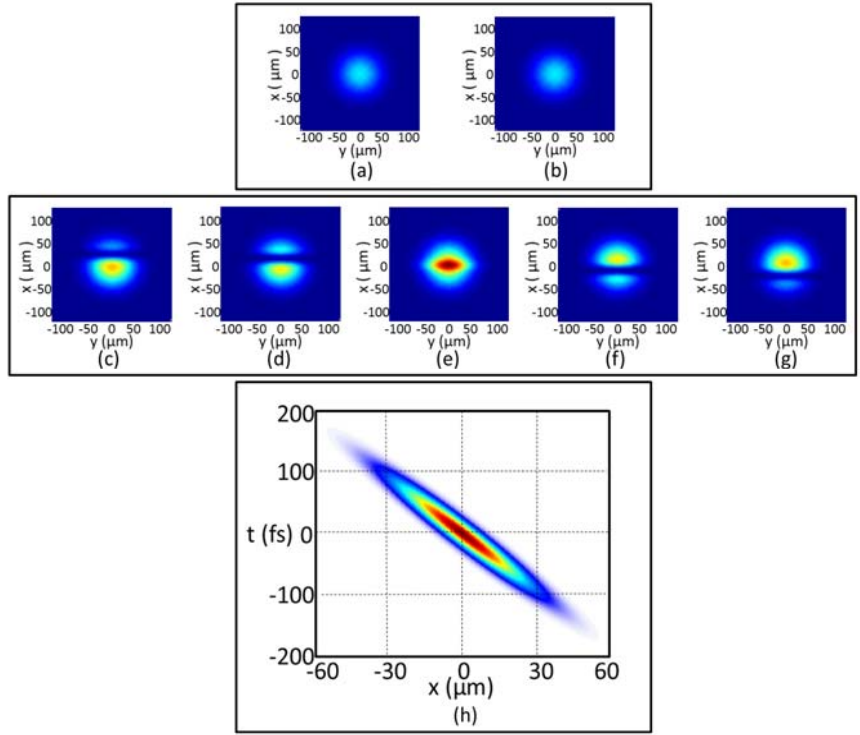

Fig. 5 Calculated intensity distribution of the focal spot in the focal plane produced in air with (a) STF and (b) CF systems. Calculated intensity distributions of the focal spots formed by superposition of the two beams are displayed in (c)-(g). The relative time delays in (c)-(g) are -66.8 fs, -33.4 fs, 0 fs, 33.4 fs, 66.8 fs, respectively. (h) displays the calculated PFT of the focal spot produced in air with the STF system.

We now theoretically examine the interference patterns formed in fused silica which will be compared with the experimental results in Fig. 4. It is noteworthy that when the pulses are focused in transparent media, spherical aberration caused by a refractive index mismatch at the interface of air and the media should be considered [23]. Such refractive index mismatch can be included in our simulation by introducing a phase aberration function $\Phi(x,y,d)$ into the light field [24, 25]. The aberration function $\Phi(x,y,d)$ is defined as follow [24]:

$$\Phi(x,y,d) = -\frac{kd}{f}\left[\sqrt{f^2 n^2 - (x^2 + y^2)} - \sqrt{f^2 - (x^2 + y^2)}\right]. \qquad (6)$$

where $n$ is the refractive index of the material, and $d$ is the nominal focusing depth of the light in fused silica. By using Eq. (9) in the appendix of reference [24], the light field in the front focal plane can be expressed as:

$$E_1'(x,y,\omega) = \exp\left[-i\hat{\Phi}(x,y,d)\right] E_1(x,y,\omega). \qquad (7)$$

The intensity distribution of the focused beam in the focal plane inside fused silica can thus be obtained by replacing the light field described by Eq. (5) with the light field described by Eq. (7) in Eqs. (1) and (2). Accordingly, the simulation results are presented in Fig. 6. Figures 6(a) and (b) present the intensity distributions of the focal spot individually produced with the CF and STF systems, respectively. In contrast, Figs. 6(c)-(g) present the interference patterns produced by simultaneously focusing the two beams with both the CF and the STF systems, in which the relative time delays between the two pulses are -66.8 fs, -33.4 fs, 0 fs, 33.4 fs, 66.8 fs, respectively. Once again, the simulation results of the interference patterns obtained with the inclusion of the spherical aberration phase are in excellent agreement with the experimental results in Figs. 4(c)-(g). The numerical simulation of the intensity distribution of spatiotemporally focused spot in *x-t* domain is shown in Fig. 6(h). The calculated degree of PFT is ~3.06 fs/μm, which is also consistent with the measured result.

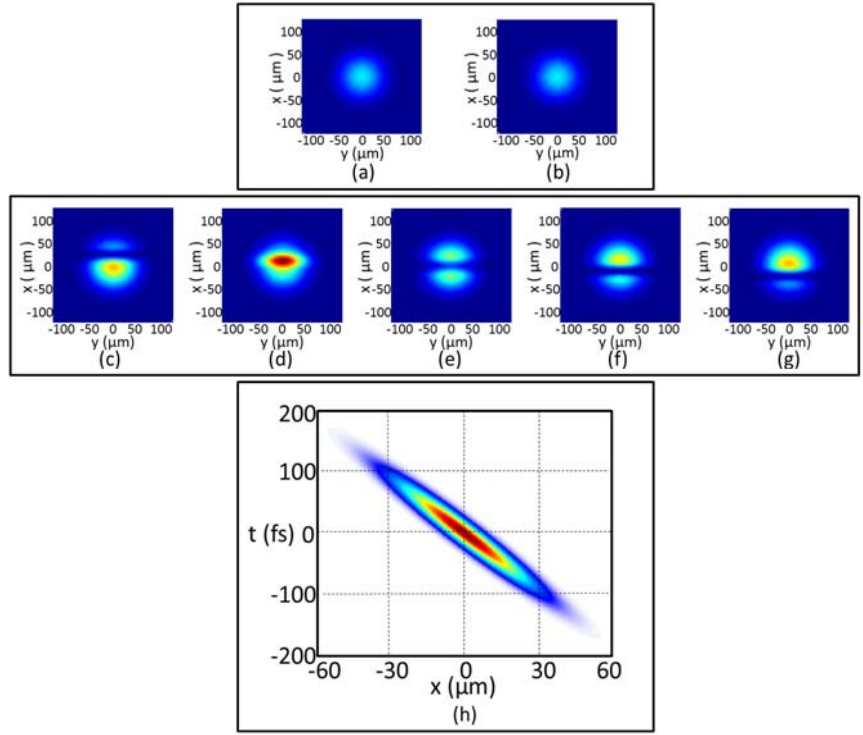

Fig. 6 Calculated intensity distribution of the focal spot in the focal plane produced in glass with (a) STF and (b) CF systems. Calculated intensity distributions of the focal spots formed by superposition of the two beams are displayed in (c)-(g). The relative time delays in (c)-(g) are -66.8 fs, -33.4 fs, 0 fs, 33.4 fs, 66.8 fs, respectively. (h) displays the calculated PFT of the focal spot produced in glass with the STF system.

We note here that in both our experiments and simulations, a focal lens of a low numerical aperture value (NA=0.007) is used. Under such condition, the influence of spherical aberration is negligible. When the numerical aperture of the focusing system (*i.e.*, an objective lens) becomes high, the influence of the spherical aberration caused by refractive index mismatch at the interface will become greater. A direct consequence is that both the focal spots produced by the STF and CF systems will be significantly distorted in terms of the amplitude and phase distributions in the focal plane. In such a case, our interferometric technique for characterization of the PFT of spatiotemporally focused pulses should be modified. For instance, one may choose a low NA value of the CF system for the reference pulse (which can be realized by reducing the beam diameter using an aperture) while maintaining a high NA value for the pulses focused by the STF system. Another interesting direction is to develop a technique that allows for direct characterization of the PFT of intense pulses in the nonlinear propagation regime. These investigations will be carried out in the future.

**5. Conclusion**

We have developed an interferometric technique for measuring the sweep velocity of a focal spot spatiotemporally focused in either air or fused silica glass. In principle, this technique can also be used for other transparent materials such as crystals and liquids. The quantitative characterization of the PFT of the spatiotemporally focused pulses can be achieved with a high precision as far as the linear propagation condition is maintained. Interaction of the spatiotemporally focused femtosecond laser pulses with transparent media has been attracting increasing attention for internal processing of transparent materials and tissue engineering, which will all benefit from the technique based on the interferometric diagnostics.

**Acknowledgments**

This work is supported by National Basic Research Program of China (No. 2014CB921300), National Natural Science Foundation of China (Nos. 61327902, 11104294, 61275205, 61405220, 61221064 and 11304330), and Shanghai "Yang Fan" program (Grant No. 14YF1406100).